\documentclass{iau} 
\usepackage{graphicx}
\usepackage{natbib}

\pubyear{2022}
\volume{369}
\jname{The Dawn of Cosmology \& Multi-Messenger Studies with Fast Radio Bursts}
\editors{E.~Keane, K.~Rajwade, A.~Fialkov \& C.~Walker, eds.}

\title[FRB emission mechanisms] 
{FRB  emission mechanisms vs. observations}

\author[Popov S.B.]   
{Popov S.B.$^{1,2}$ }

\affiliation{$^1$Sternberg Astronomical Institute, 119234, Universitetski pr. 13, Moscow, Russia \\ [\affilskip]
$^2$Department of Physics, Lomonosov Moscow State University, 119991, Moscow, Russia}

\begin{document}

\maketitle

\begin{abstract}
 Presently, it is broadly assumed that fast radio bursts (FRBs) are sources of coherent emission powered by the magnetic energy release in magnetars. However, the exact emission mechanism is not known, yet. Two main frameworks exist: magnetospheric emission and radiation from external relativistic shocks. In this brief review, I describe basics of both approaches and discuss how they are probed in modern observations.
\keywords{radiation mechanisms: nonthermal, stars: neutron}
\end{abstract}

\firstsection 
              
\section{Introduction. Where do we stand?}

After 15 years of studies, we know quite a lot about FRBs (see e.g., a recent broad review by \citealt{caleb21}). 
Regarding the topic of this brief review, I want to underline three main items which can be considered as the base of the present day 'standard' FRB model:

\begin{itemize}
\item Coherent emission;
\item Magnetic energy release;
\item Magnetars as the sources.
\end{itemize}
In the following three paragraphs I briefly comment on each of them.

 Already in the discovery paper \citep{lorimer2007} the authors demonstrated that the brightness temperature might be $\sim 10^{34}$~K and thus, a coherent emission mechanism is necessary. 
All later observations only supported this conclusion.
 So, from the very beginning development of all reliable mechanisms is focused on a coherent emission generation. 

 Since 2007 dozens of various types of sources and emission mechanisms have been proposed (see \citealt{xiao21} and references therein). Still, relatively quickly two of them started to attract the main attention: magnetars and 'pulsars on steroids'. In the former, magnetic energy $E_\mathrm{mag}$ is released, in the latter -- rotational energy, $E_\mathrm{rot}$. 
 Potentially,  $E_\mathrm{rot}=\frac12 I \omega^2\sim \mathrm{few} \times 10^{52} p_{-3}^{-2}$~erg is significantly larger than $E_\mathrm{mag}\sim \mathrm{few} \times  10^{47} B_{15}^2$~erg. 
 Still, step by step pulsar-like mechanisms started to face difficulties. 
 In particular, rotation powered mechanisms hardly can produce very energetic bursts, or revealed parameters imply a rapid evolution of the source, which is not observed \citep{lyutikov2017}.
 Presently,  most of prominent approaches involve just magnetic energy release.

 Finally, after the celebrated discovery in April 2020 of simultaneous radio and high energy bursts from the Galactic source SGR 1935+2154  (see a brief review in Sec.~6 of \citealt{nicastro21}, references to the original papers are given below), the original proposal that FRBs are related to flashes of magnetars \citep{pp2007} got a strong support. Recently (October 2022), a few more coincident radio and high energy events were detected from the same source \citep{15681, 15682, 15686, 15707, 15708}. Although this does not prove that all FRBs are due to a magnetar activity, magnetars seem to be the best site to produce FRBs.

 Below, I shall discuss only the mechanisms related to a coherent emission generation powered by the magnetic energy release in magnetars.


\section{Emission mechanisms}

 The most popular approaches to describe radio emission of FRBs can be divided in the two families of models (see an excellent brief review by \citealt{zhang2020}). The first one deals just with magnetospheric processes.
In the second approach, pioneered by \cite{lyub2014}, radio emission is generated in relativistic shocks outside the magnetosphere.
Both scenarios (below I dub them as  magnetospheric and external, correspondingly) include an initial perturbation in the magnetosphere accompanied by significant energy release most of which might be emitted at high energies. The radio emission is generated either in the magnetosphere and then propagates through it outwards, or a part of the energy of the flare is reprocessed in an external medium producing radio waves. 

It is not for granted that all FRBs are caused by a single emission mechanism. 
Still, different authors mostly try to explain in one model the whole set of the main parameters observed in various sources (repeating or not). The entire list of these parameters can be very long, just to name a few:

\begin{itemize}
    \item Range of luminosities;
    \item Various morphological types; 
    \item Duration;
    \item Rapid variations of the flux;
    \item Brightness temperature;
    \item Repetitions (sometimes with a very high rate);
    \item Very high degree of polarization in many cases;
    \item Variations of polarization from burst to burst in repeaters; 
    \item Relatively narrow spectra (sometimes variable from burst to burst in the case of  repeating sources);
    \item Frequency drift ('sad trombone' effect, etc.).  
\end{itemize}

 In the following two subsections I present a sketch of both scenarios, and then in Sec.~3 briefly discuss correspondence between theoretical predictions and observational data.


\subsection{Synchrotron maser models}

The idea of a coherent emission generated at relativistic shocks has been discussed in relation to astrophysical applications long before the discovery of FRBs \citep{hoshino}. 
During the last $\sim8$~years this model (including a synchrotron maser emission) is actively developed by many groups all over the world to explain properties of FRBs (see a comprehensive review by \citealt{lyub2021}). 

 In this framework an ejecta (or a 'magnetic piston' -- an electromagnetic pulse dominated by the magnetic field) formed in the magnetar's magnetosphere propagates outwards with an ultra-relativistic speed and interacts with the surrounding medium forming a relativistic shock. 
For example, in the model by \cite{beloborodov20} a magnetar via reconnection produces a plasmoid which moves with a Lorentz factor $\Gamma_\mathrm{pl} \gtrsim 10^5$. It starts to interact with the magnetar's cold relativistic wind with $\Gamma_\mathrm{w}\sim(10-20)$. This results in the formation of a shock wave with $\Gamma\gtrsim 10^3$. In the shock wave downstream the conditions corresponding to a synchrotron maser can be fulfilled. Then, at a typical distance $\sim10^{14}$~cm from the magnetar  a millisecond-long radio burst with a typical frequency $\sim$GHz is produced. 
 
 In order to demonstrate a different approach in the same framework, we can look at the model by \cite{khan21}. These authors presented a scenario where a radio burst is generated at the reverse shock which appears after interaction of a powerful magnetar flare with a magnetar's wind termination shock. For the terminal shock distance $\lesssim 10^{15}$~cm the characteristic frequency of emission is about 1 GHz. In this model  only $\sim$10\% of magnetars are in the evolutionary
phase suitable for the production of FRBs and just a small fraction ($\sim 10^{-4}$) of their flares produce FRBs. This helps to resolve several issues known from observations.

\subsection{Magnetospheric emission models}

 Some early models of FRBs utilised magnetospheric processes to generate emission due to rotational energy losses (e.g., \citealt{katz14}). However, in this subsection I discuss only scenarios based on magnetic energy dissipation. Three scenarios will be mentioned, however there are many more magnetospheric models.
There detailed discussion requires a special review. 

 In the first one proposed by \cite{lu20}, disturbances are propagating along opened field lines, which finally results in generation of radio emission (while a high energy flare is generated by plasma trapped in the closed lines part of the magnetosphere). Conversion of Alfven waves into radio waves happens at the charge starvation region at $\sim10^8$~cm.

Quite a different scenario was proposed by \cite{lyub2020}. 
In this model a magnetic pulse is generated after a magnetar flare. It propagates outside the light cylinder till it hits the current sheet.
This results in reconnection of field lines, which finally leads to radio emission.
In some sense, this is a 'hybrid' model (in some respect similar to the external scenario), but I prefer to put it together with the magnetospherics ones, as it involves just the structure which is intrinsic for a magnetosphere of a rotating neutron star.

Finally, \cite{lyutikov2021} used a free electron laser concept in application to radio emission of highly magnetized neutron stars. In this case a relativistic beam of electrons, generated due to a reconnection event,  must propagate through periodically varying magnetic field. In the case of neutron stars such a field structure can appear e.g., due to the firehose instability in the magnetosphere. Then,  bunches of particles are formed in the beam, and particles in a bunch emit coherently producing a FRB. 

Notice, that in these three models energy is carried towards the site of radio emission by three different subjects: MHD waves, a magnetic pulse pushing plasma, and a particle beam. 

In the next section I summarize how successful are the two main frameworks -- magnetospheric and external, -- in fitting the  present observational data.


\section{Observations and theory}

 Despite a very rich phenomenology, the exact emission mechanism for FBRs remains unknown, yet. 
 In many respects, this situation resembles the one with radio pulsars (e.g., \citealt{beskin18}). 
Although, thousands of pulsars are known and many of them are observed in all wavelengths -- from radio to gamma, -- there is no consensus about important issues regarding generation of their radiation (again, from radio to gamma), despite 55 years of studies. Hopefully, for the FRBs we can find a solution faster.

 Here I select just eight points, which are briefly reviewed below:

\begin{itemize}
\item Polarization;
\item Periodicity;
\item Narrow spectra;
\item Frequency drift;
\item Rapid variability; 
\item Repeaters vs. non-repeaters;
\item Burst rate and total energetics;
\item Bursts from SGR 1935+2154.
\end{itemize}

 A review of polarization properties of FRBs can be found e.g., in Sec~5.9 of \cite{caleb21}.
Polarization is measured for more than 20 FRBs (including several repeaters). 
In a few cases (again, including repeaters) the measured degree of linear polarization is compatible with 100\%.

An important result was presented by \cite{luo20}. 
Analysing 15 bursts from  FRB 180301 these authors discovered significant changes in 
polarization properties from burst to burst. This feature does not easily fit external models, but can be  explained in the magnetospheric scenario, naturally.

Oppositely, observations of FRB 121102 \citep{michilli18} demonstrated a very stable polarization angle and nearly 100\% linear polarization for all bursts. Such stability is not expected for magnetospheric emission.  
However, \cite{lyub2020} noticed, that it is non-trivial to explain 100\% polarization in the framework of relativistic shock emission. 

A unique result was obtained by \cite{andersen22}. 
A non-repeating source FRB 20191221A demonstrated a 3-second long burst with a complicated pulse structure.
Analysis allowed to derive a periodicity with the period $\sim217$~msec at a $\sim6.5\, \sigma$ level.
As it was noted by the authors, such periodicity can be easier explained in magnetospheric models than in the external framework.

Much effort has been expended to detect FRBs simultaneously at different radio frequencies (from  $\sim 100$ MHz up to several GHz) in order to measure spectral parameters in a wide range.
However, all such attempts (especially intensive for repeaters, like FRB 121102 and R3) were so far unsuccessful.
For example, \cite{pastor21} observed FRB 20180916B with LOFAR (110-190 MHz) and Apertif/Westerbork (1.2-1.5 GHz). 
Many detections have been made by each instrument, but not a single one simultaneously at high and low frequencies. Similar results were obtained for several sources by different authors. Simultaneous detections are  done only at close frequencies, e.g. like in the case of FRB 121102 at 1.4 and 3 GHz (but without detections at 4.85 and 15 GHz), see \cite{law17}. 
This points to narrow spectra of FRBs.

An extreme example is given by \cite{kumar21}.
Using the 64-meter Parkes telescope these authors detected a very short (1 msec) burst from the repeater FRB 20190711A. The peculiarity of the result is related to an extremely narrow band in which the source was detected: just 65 MHz at $\sim1.4$~GHz.  It is not clear, if this feature is directly related to the emission mechanism, but if it does -- then it can be a challenge for many  models. 
Specifically accounting for the fact, that this repeating source shows different spectral properties for different bursts. 

Generally, explanation of the variety of spectral properties (including recent detection of a radio burst from SGR 1935+2154 at 400 MHz and 5.6 GHz simultaneously with high-energy detection, see \citealt{15697}) and their variability in repeating sources
might be a good tool to probe theoretical models, especially if the aim is to explain all sources in one framework. 

 FRBs can also have unusual dynamical spectra: spectral properties can change during a burst or from one sub-burst no another. 
The most known is so-called 'sad trombone' effect, when the frequency of emission is decreasing during a burst \citep{gajjar18}. 

 The 'sad trombone' effect can be naturally explained in the magnetospheric approach (e.g., \citealt{lyutikov2020}).
As emitting particles move in the magnetosphere outwards, the frequency of their radiation goes down because the emission site is situated at larger and larger distance (i.e., in lower magnetic field) for later moments of time.
In the external framework the frequency down drift  can be explained, too, e.g., via a deceleration of the shock (see \citealt{metzger22} and references therein). 

 Often, subsequent sub-bursts have frequencies shifted down, somehow similar to the 'sad trombone' effect (see examples in \citealt{gajjar18}). However, sometimes opposite behavior is observed: sub-bursts demonstrate frequency increase \citep{zhou22}. Explanation of such behavior in one framework together with 'sad trombone' and other peculiarities can be difficult for both concurrent approaches. 

Another important observational feature is related to the shortest time scale of variability found in FRBs. At the moment, the record probably belongs to  FRB 20180916B.  \cite{nimmo2021} reported observations of sub-pulses at the time scale about few $\mu$s. Such short time scales are more naturally explained in the magnetospheric scenario. However, small scale inhomogeneities in the external medium, where the shock propagates, can help to obtain $\sim\mu$s variability also in synchrotron maser models. 
Larger statistics of short flux variations in repeating sources potentially can help to distinguish between predictions of the two frameworks.  

One of the most disputable questions in phenomenology of FRBs is related to a possible dichotomy between repeaters and non-repeaters. Extensive comparison was made by \cite{pleunis21}. These authors used 61 bursts from 18 repeaters plus 474 events from non-repeaters, all detected by CHIME.  Clearly, two groups of bursts are well-separated in terms of spectral and temporal parameters (with a possible addition of the third group of one-off events from not-yet-identified repeaters). Repeaters have larger bursts duration and more narrow band width.  

Note that many (indirect) arguments in favour of the magnetospheric framework mentioned above come from observations of repeaters. Then, it is tempting to hypothesize that non-repeaters (or one-off events)  are due to the competitive mechanism -- synchroton maser radiation of relativistic shocks. However, this idea can be criticized. One-off events have very narrow distributions in duration and band width (see \citealt{pleunis21}). This is not naturally expected in the external framework due to either different properties of the external medium in different sources, or due to different parameters of the magnetar wind, if its properties are important.  

 Some problems for the 'standard' model of FRBs can appear due to observations of 'burst storms' from several sources \citep{xu22, nimmo22, zhou22}. If we assume that the ratio of radio luminosity to the total energy release in a flare is $\sim10^{-5}$, as it is suggested from observations of the Galactic source (see below), then the total magnetic energy losses can be $\gtrsim10^{46}$~erg just in few days. Total magnetic energy available for powering a magnetar's flares hardly can exceed $10^{48}$~ergs. Thus, such a strong activity might be a very short, $\sim 10$~yrs, episode in a life of the magnetar. Appearance of this short evolutionary stage requires a satisfactory explanation. 

 In addition, the huge energy release during such active periods might significantly modify properties of the external medium. This can be important for some external models. So, it is promising to look for systematic differences between bursts emitted during periods of strong activity and bursts from more quiet sources. 

For several years one of the key arguments against the hypothesis that magnetars are the sources of FRBs was a simple question: why we do not see flares from Galactic objects?  Finally, on April 28, 2020 simultaneous bursts in radio and in X/$\gamma$-rays were detected from the Galactic source SGR 1935+2154 \citep{chime, stare, mereg20, ridnaia, agile, hmxt}. 

As this is the very first detection of a high-energy counterpart of an FRB,  now we have the first estimate of the fraction of the total energy of a flare emitted in radio: $E_\mathrm{radio}/E_\mathrm{total}\sim 10^{-5}$. This value can be explained in both frameworks \citep{lyutikov2021, yang21, yuan20}.

 An intriguing feature of FRB 200428 is related to the delay between radio and high energy emission: at low frequencies peaks of the bursts are emitted a few millisecond earlier than in the opposite part of the electromagnetic spectrum \citep{mereg20, hmxt}. Curiously, this peculiarity also can be explained in both frameworks \citep{lyutikov2021, yuan20}. 
New detections of simultaneous bursts from this source in October 2022 (see \citealt{15681, 15682, 15686, 15707, 15708}) can provide new information on this issue, but at the time of writing of this review details are not published, yet.

In addition to the importance of observations of simultaneous bursts, there is a question why among hundreds of high energy and radio bursts detected from SGR 1935+2154 just a few come together 
(see e.g., \citealt{lin20, younes2020}). The first FRB-like event (coincident with an X-ray flare) from this source had peculiar properties in the high energy range \citep{mereg20, ridnaia, agile, hmxt}. But at least the first of the October bursts observed simultaneously by radio telescopes and gamma-ray detectors has properties much closer to the rest of high energy events \citep{15686}. This variety of properties must be explained in a successful model.



\section{Future outlook}

 To conclude, magnetars seem to be the sources of (at least, most of) FRBs. The exact mechanism of emission is unknown, but two main frameworks are formulated: magnetospheric processes and coherent (e.g., maser) emission from external relativistic shocks. Up to now, plethora of observational data cannot exclude either of the main scenarios.

 Some progress in disentangling between different models of FRB emission may be achieved due to variety of properties of  medium in vicinity of the sources. 
 If we accept the magnetar model, then there is a question regarding the origin of these magnetars.
 Already now, we can formulate some ideas about it, as FRBs are detected at very different sites.
For $\sim20$ sources host galaxies are identified \citep{heintz20}. 
 The prolific repeater FRB 121102 is known to be situated in a starforming region in a dwarf galaxy \citep{kokubo17}. 
Similar situations are realized for several other sources. On other hand, some sources are localized in galaxies with low starformation rate. An extreme example of such situation is the source FRB 20200120E which is located in a globular cluster of  the M81 galaxy \citep{kirsten22}. 
It is natural to assume that magnetars in these cases have different origin. In regions of active starformation they might mostly come from a normal core collapse. In globular clusters and early type galaxies they, most probably, originate from coalescence of compact objects or accretion induced collapse. 
 Thus, we can expect quite different environment in these cases.
 Finally, if periodicity of FRB 121102 \citep{rajwade20} and FRB 180916 (R3) \citep{amiri20} is due to the orbital motion in a binary, then magnetars there are embedded in a dense stellar wind of the massive companion (see e.g., \citealt{barkov21} and references therein). Which again suggests a specific type of surrounding medium.

 Properties of the environment are important in several models involving external shocks, mentioned  above. 
 If parameters of the surrounding medium are not determined solely by the magnetar wind, then we can expect significant differences in the conditions at the shock, which might manifest themselves in properties of FRB emission.
 For example, pressure in the external medium is important in models by \cite{lyub2014} and \cite{khan21}. 
Then we can expect that in significantly different environments properties of the observed FRBs are also systematically different.

Many hopes are bind to observations of Galactic magnetars. 
While perspectives to detect counterparts of extragalactic FRBs are very vague, in our own Galaxy we have
now three detections of simultaneous high-energy and radio bursts from SGR 1935+2154.\footnote{On October 14, 2022 the second simultaneous event was detected, see \citealt{15681, 15682, 15686, 15697}. And then, the third event, see \citealt{15707, 15708}.} 
Detailed observations of relatively near-by sources might provide insight into the emission mechanism. 
It is particularly interesting to measure precisely if there is a time delay between radio and high-energy signals in all cases. E.g., this feature was discussed by \cite{lyutikov2021} in his free electron laser approach. 

Another feature, which can help to shed light on the emission mechanism, is spin periodicity, especially
in a series of consecutive bursts, like in the case of FRB 20191221A \citep{andersen22}.
A solid proof of spin period origin of such events can be a strong argument in favour of the magnetospheric origin of radio emission. 

And, of course, it well can be that some serendipitous discoveries (e.g., high energy counterparts of extragalactic FRBs) will finally uncover the emission mechanism of FRBs.

\acknowledgements
I acknowledge support from  the Russian Science Foundation (grant 21-12-00141).

Also I thank dr. D.~Khangulyan for many useful comments on the manuscript.


\def\apj{{ApJ}}  
\def\apjs{{ApJS}}
\def\nat{{Nature}}    
\def\jgr{{JGR}}    
\def\apjl{{ApJ Letters}}    
\def\aap{{A\&A}}   
\def\mnras{{MNRAS}}
\def\aj{{AJ}}
\let\mnrasl=\mnras


\end{document}